\documentclass[twocolumn,secnumarabic,amssymb, nobibnotes, aps, prl, superscriptaddress]{revtex4-2}

\usepackage{graphicx}
\usepackage{dcolumn}
\usepackage{bm}
\usepackage{amsmath}
\usepackage{enumerate}
\usepackage{setspace}
\usepackage{dsfont}
\usepackage{subfigure}
\usepackage{multirow}
\usepackage{indentfirst} 
\usepackage {mathrsfs}
\usepackage{color}

\usepackage{lineno}

\usepackage{makecell}   
\usepackage{multirow}	

\usepackage{threeparttable}  
\usepackage[pdfstartview=FitH,
CJKbookmarks=true,
bookmarksnumbered=true,
bookmarksopen=true,
colorlinks, 
pdfborder=001,
linkcolor=blue,
anchorcolor=blue,
citecolor=blue
]{hyperref}

\begin{document}
	

\title{Dual-comb-enhanced microwave clock synchronization over commercial fiber}%

\author{Ziyang Chen}%
\email[E-mail: ]{chenziyang@pku.edu.cn}
\affiliation{State Key Laboratory of Advanced Optical Communication Systems and Networks, School of Electronics, and Center for Quantum Information Technology, Peking University, Beijing 100871, China}

\author{Dongrui Yu}%
\affiliation{State Key Laboratory of Advanced Optical Communication Systems and Networks, School of Electronics, and Center for Quantum Information Technology, Peking University, Beijing 100871, China}

\author{Ganbin Lu}%
\affiliation{State Key Laboratory of Information Photonics and Optical Communications, Beijing University of Posts and Telecommunications, Beijing 100876, China}

\author{Yufei Zhang}%
\affiliation{State Key Laboratory of Advanced Optical Communication Systems and Networks, School of Electronics, and Center for Quantum Information Technology, Peking University, Beijing 100871, China}

%
%
%
%
\author{Song Yu}%
\affiliation{State Key Laboratory of Information Photonics and Optical Communications, Beijing University of Posts and Telecommunications, Beijing 100876, China}

\author{Bin Luo}%
\email[E-mail: ]{luobin@bupt.edu.cn}
\affiliation{State Key Laboratory of Information Photonics and Optical Communications, Beijing University of Posts and Telecommunications, Beijing 100876, China}

\author{Hong Guo}%
\email[E-mail: ]{hongguo@pku.edu.cn}
\affiliation{State Key Laboratory of Advanced Optical Communication Systems and Networks, School of Electronics, and Center for Quantum Information Technology, Peking University, Beijing 100871, China}

\date{\today}%

\begin{abstract} 
The large-scale clock network is the key ingredient to obtain high precision in many scenarios, from fundamental research to cutting-edge applications. 
The advantage of the time synchronization among microwave clocks is their cost, size, and accessibility.
Here, we demonstrate a femtosecond-level time synchronization of microwave clocks through a commercial link of 205.86 km via dual-comb-enhanced optical two-way time transfer, which achieves a 6.23-fs residual time deviation between synchronized timescales at 1 s and an instability below $6 \times {10^{ - 18}}$ at 10,000 s. Further, the high-precision time synchronization of microwave clocks significantly enhances the probe ability of subtle reciprocity changes of fiber to the sub-picosecond level.
This work provides a path toward secure fiber time-frequency networks to support future microwave-clock-based precise timing and sensing systems.
\end{abstract}

\maketitle

\section{Introduction}

High-precision clock networks offer the fundamental technical foundation for numerous scenarios, including positioning, navigation, timing~\cite{J.Geod.89.607.2015,Metrologia.48.S219.2011,J.Geod.83.191.2009}, sensitive probes of environmental fluctuations~\cite{Science.376.874.2022,Science.361.486.2018,Optica.5.893.2018}, and fundamental physics testing~\cite{Nature.591.564.2021,Phys.Rev.Lett.118.221102.2017,Nat.Commun.7.12443.2016,Nat.Phys.10.933.2014}. The enormous benefits are unleashed through the full interconnection between any two nodes in a hybrid network of optical clocks and microwave clocks. In particular, the
time-frequency transfer over both 
fiber and free-space optical links
has seen tremendous progress~\cite{APL.Photonics.9.016112.2024,Nature.618.721.2023,Nature.610.661.2022,Nat.Commun.13.212.2022,Phys.Rev.Lett.128.020801.2022,Nat.Commun.12.515.2021,Optica.8.471.2021,APL.Photonics.5.076113.2020,Nat.Commun.10.1819.2019,Phys.Rev.Lett.120.050801.2018,Phys.Rev.X.6.021016.2016,Optica.3.441.2016,Applied.Physics.Letters.109.151104.2016,Nature.Photon.7.434.2013,Phys.Rev.Lett.111.110801.2013,Science.336.441.2012,Nature.Photon.4.716.2010,Nature.Photon.2.733.2008,Phys.Rev.Research.6.023005.2024,Appl.Phys.Lett.110.221101.2017}. The ultra-high precision frequency comparison has been achieved using well-established optical frequency transfer techniques by directly disseminating stable frequency signals~\cite{Nat.Commun.13.212.2022,Phys.Rev.Lett.128.020801.2022,Nat.Commun.12.515.2021,Phys.Rev.Lett.111.110801.2013,Science.336.441.2012}, whereas the timescale comparison requires an exchange of
a series of labeled optical pulses generated by frequency standards~\cite{APL.Photonics.9.016112.2024,Nature.618.721.2023,Nature.610.661.2022,Optica.8.471.2021,APL.Photonics.5.076113.2020,Nat.Commun.10.1819.2019,Phys.Rev.Lett.120.050801.2018,Phys.Rev.X.6.021016.2016,Optica.3.441.2016,Applied.Physics.Letters.109.151104.2016,Nature.Photon.7.434.2013}. 
Since the optical two-way time transfer (OTWTT) method uses the reciprocity of bi-directional light travel to eliminate most of the common-mode noises and the link drift~\cite{Metrologia.50.133.2013,Metrologia.49.772.2012}, the measurement of optical pulses has become the key to improving the precision for time transfer.

Remarkably, in recent years, time transfer has been significantly improved by using optical measurement methods, specifically linear optical sampling (LOS)~\cite{Opt.Lett.34.2153.2009}, to overcome the limitation of the picosecond-level uncertainty caused by the photoelectronic detection of pulses~\cite{Nature.610.661.2022,Optica.8.471.2021,APL.Photonics.5.076113.2020,Nat.Commun.10.1819.2019,Phys.Rev.Lett.120.050801.2018,Phys.Rev.X.6.021016.2016,Optica.3.441.2016,Applied.Physics.Letters.109.151104.2016,Nature.Photon.7.434.2013,Photon.Res.11.2222.2023,Review.of.Scientific.Instruments.94.085105.2023}. State-of-the-art time-programmable frequency comb technology has even pushed the time synchronization to the quantum-limited region~\cite{APL.Photonics.9.016112.2024,Nature.618.721.2023}. Despite great improvement of the synchronization capability between optical clocks~\cite{Nature.618.721.2023,Nature.610.661.2022,Phys.Rev.X.6.021016.2016} and between optical and microwave clocks~\cite{Optica.3.441.2016,Applied.Physics.Letters.109.151104.2016}, the reported time synchronization between remote microwave clocks remains limited by the precision of photodetectors and time-interval counters (TICs)~\cite{Metrologia.41.17.2004}, so the full potential of high-precision time-frequency networks has not been realized.

In the time synchronization of two microwave clocks, the electrical measurement has not been replaced by the optical approach because a phase coherence of independent optical frequency combs (called combs for simplicity) at both local and remote sites, which is a necessary requirement for performing stable and low-jitter interference, is challenging to establish. Specifically, microwave-to-microwave synchronization is difficult because both reference and synchronized clocks have greater phase noise than optical clocks and the pulse coherence is reduced by propagation-induced pulse impairments. It was challenging to consider not only the impact of inferior reference sources on synchronization but also the effects of fiber on long-distance optical frequency comb transmission. These factors collectively affect the precision sharing of remote microwave clocks.

To overcome the limitations mentioned above, here, we propose a tight microwave-clock time synchronization scheme based on the dual-comb-enhanced OTWTT and provide a demonstration over 205.86 km commercial fiber. 
We achieved a 6.23-fs residual time deviation between synchronized timescales at 1 s and an instability below $6 \times {10^{ - 18}}$ at 10,000 s due to a series of breakthroughs, including high-precision locking, LOS, dispersion compensation, robust timing discrimination, and signal processing. The intensity-encoded comb pulses were measured using a Kalman filter–assisted LOS, where the centroids of reference and target interferograms at each site served as the start and stop times of the time interval, to reduce the impact of the intensity noise and pulse distortion on the peak measurements.
We used the phase time of the pulse envelope instead of the carrier phase for time synchronization~\cite{Phys.Rev.Lett.120.050801.2018} to remove the limit of the carrier envelope offset (CEO) frequency, since the stable carrier phase is vulnerable to fluctuations in independent sources and fiber transmission. Dispersion compensation, fully automatic polarization control, and real-time feedback are essential to guarantee the mutual optical coherence.

In contrast to the majority of reported works~\cite{Nature.618.721.2023,Nature.610.661.2022,Optica.3.441.2016,Nat.Commun.10.1819.2019}, our research specifically addressed the long-haul time synchronization scenario by transmitting optical frequency combs through optical fibers, under the conditions of utilizing microwave clocks with inferior noise characteristics. Although our approach for time synchronization using microwave clocks as reference sources achieves slightly lower precision compared to time synchronization schemes using optical clocks, we believe that the improvement in microwave clock synchronization capabilities has greatly expanded the application range of the entire high-precision clock networks.

An interesting extension of the fiber-based time synchronization is the perception of subtle reciprocity changes of the fiber, which enhances the ability against potential delay attacks~\cite{IEEE.Trans.Instrum.Meas.72.1.2023,Nat.Phys.16.848.2020,IEEE.J.Sel.Top.Signal.Process.12.749.2018}. We achieved sub-picosecond (ps)-level sensing of asymmetric delay attacks within 200 kilometers, which significantly surpasses the capabilities of the existing TIC measurement instruments of a sub-nanosecond level~\cite{IEEE.Trans.Instrum.Meas.72.1.2023}.

\section{Concept}

The main task of this work was to achieve synchronization of optical pulses carrying time information from two microwave clocks, one located at the master site (site A) and the other at the synchronized site (site B). Site A and site B are adjacent to each other and located at one laboratory with independent rubidium clocks to easily quantify the residual transfer noise. Note that the microwave clock in this paper refers to both the microwave oscillator (a rubidium clock) and the optical ``clock" comb that was being synchronized. The overview of the microwave-clock-based time synchronization via fiber is given below, and the specific experimental layout will be introduced in the next section.

(1) Creating the local timescales.  At each site, a local timescale is created by phase-locking the repetition frequency of the ${f_{\rm{r}}}$ comb to the local microwave clock, which served as the clock comb. Then, a ${f_{\rm{r}}} + \Delta {f_{\rm{r}}}$ comb, which is also locked to the local clock, output optical pulses with a repetition frequency offset by $\Delta {f_{\rm{r}}}$ relative to the clock comb, which serves as the local oscillator (LO) comb to sample clock combs for performing LOS.

(2) Generating band-limited pulses. The international-telecommunication-union-standard (ITU) dense wavelength division multiplexing (DWDM) module is applied to the output of each comb to generate band-limited pulses with different spectra. The LO comb is filtered to generate two outputs with a bandwidth of $B_0$, which are centered at ${\lambda _1}$ and ${\lambda _2}$. The clock comb is then filtered to generate an output centered at ${\lambda _1}$ (for site A) or ${\lambda _2}$ (for site B).

(3) Elimination of the ambiguity. Since directly using the LOS for synchronization would introduce the ambiguity of ${1 \mathord{\left/{\vphantom {1 {{f_{\rm{r}}}}}} \right.\kern-\nulldelimiterspace} {{f_{\rm{r}}}}}$, which would cause cycle slips, we provide an unambiguous synchronization method by encoding timemarkers on the optical pulse using Mach--Zehnder modulator (MZM) (Methods). For each site, the output after the intensity modulator is separated into two parts. One part is sent to the link for transmission; the other aims to generate the local reference interferogram.

(4) Dispersion compensation. Link disersion is compensated by using dispersion compensating fibers (DCFs) at the repeater station and both receiver sites. Especially, a small amount of DCF at each receiver site is used to compensate for the residual dispersion in a fine-tuned manner. For more precise dispersion compensation, techniques such as liquid crystal modulation could be used~\cite{Opt.Lett.23.283.1998}.

(5) Dual comb–based LOS measurement.  At each site, two cross-correlations are generated. The reference interferogram is generated by the cross-correlation between the LO comb and the local clock comb. The target interferogram is generated by the cross-correlation between the LO comb and the incoming clock comb from the remote site, which correspond to the “time-stretched" start and stop signals of the measuring time intervals, respectively.  The equivalent timescale stretching parameter is
obtained as $\alpha  = {{{f_{\rm{r}}}} \mathord{\left/{\vphantom {{{f_{\rm{r}}}} {\Delta {f_{\rm{r}}}}}} \right. \kern-\nulldelimiterspace} {\Delta {f_{\rm{r}}}}}$. Then, an analog to-digital converter (ADC) is implemented to digitize the train of interferograms. The data is subsequently transmitted to a field programmable gate array (FPGA) controller and synchronized with the local clock for data processing.

(6) Timing discrimination. The centroid of the interferogram is used to discriminate the temporal position of the optical pulses, defined by
\begin{equation}
	{t_i} = \frac{{\int {dt} {A_i}\left( t \right) \times t}}{{\int {dt} {A_i}\left( t \right)}},
	\label{centroidCalculation}
\end{equation}
for the $i$-th labeled interferogram. Here, ${{A_i}\left( t \right)}$ is the envelope extracted by the Hilbert transformation, since it is not sensitive to the pulse shape and can smooth out the intensity noise superimposed on optical pulses. Then, the time difference is obtained by comparing the centroid position of two interferograms.

(7) Kalman filter–assisted LOS measurement. The Kalman filter is then utilized to suppress measurement noise.

(8) In-loop time offset calculation. The real-time timing discrimination of both sites and the calculation of the in-loop time offset $\Delta {t_{_{{\rm{AB}}}}}$ are performed in the two FPGA controllers. First, at both sites, the respective time differences under a laboratory timescale $t_{\rm{A}}^{{\rm{lab}}}$ and $t_{\rm{B}}^{{\rm{lab}}}$ are obtained by locally comparing the centroids of the labeled reference and target interferograms using the time-difference calculation algorithm. The effective (actual) time differences
at both sites are ${t_{\rm{A}}}={{t_{\rm{A}}^{{\rm{lab}}}} \mathord{\left/	{\vphantom {{t_{\rm{A}}^{{\rm{lab}}}} \alpha }} \right.		\kern-\nulldelimiterspace} \alpha }$ and ${t_{\rm{B}}}={{t_{\rm{B}}^{{\rm{lab}}}} \mathord{\left/	{\vphantom {{t_{\rm{B}}^{{\rm{lab}}}} \alpha }} \right.		\kern-\nulldelimiterspace} \alpha }$. Then, the local generated time difference ${t_{\rm{A}}}$ at site A is transmitted to site B using a modulated laser over the DWDM communication link. Combined with the local calculated time information ${t_{\rm{B}}}$, the in-loop time offset $\Delta {t_{_{{\rm{AB}}}}}$ is calculated in the FPGA controller of site B.

(9) Remote clock synchronization. The time offset calculation, which acts as the error signal, is fed into the digital-feedback system that contained a proportional integral loop filter and a direct digital synthesizer (DDS). The synchronization is achieved by adjusting the repetition frequency of the clock comb at site B.

(10) Out-of-loop verification.
To verify the time offset of the clocks after synchronization, the out-of-loop measurement of the time offset $\Delta {T_{{\rm{AB}}}}$, which is independent
from the in-loop calculated time offset $\Delta {t_{_{{\rm{AB}}}}}$, is performed. An output on both combs at each site is filtered to generate the out-of-loop reference interferogram that is independent from the in-loop signals to verify the synchronization results.

\begin{figure*}[t]
	\centering
	\includegraphics[width=0.9 \linewidth]{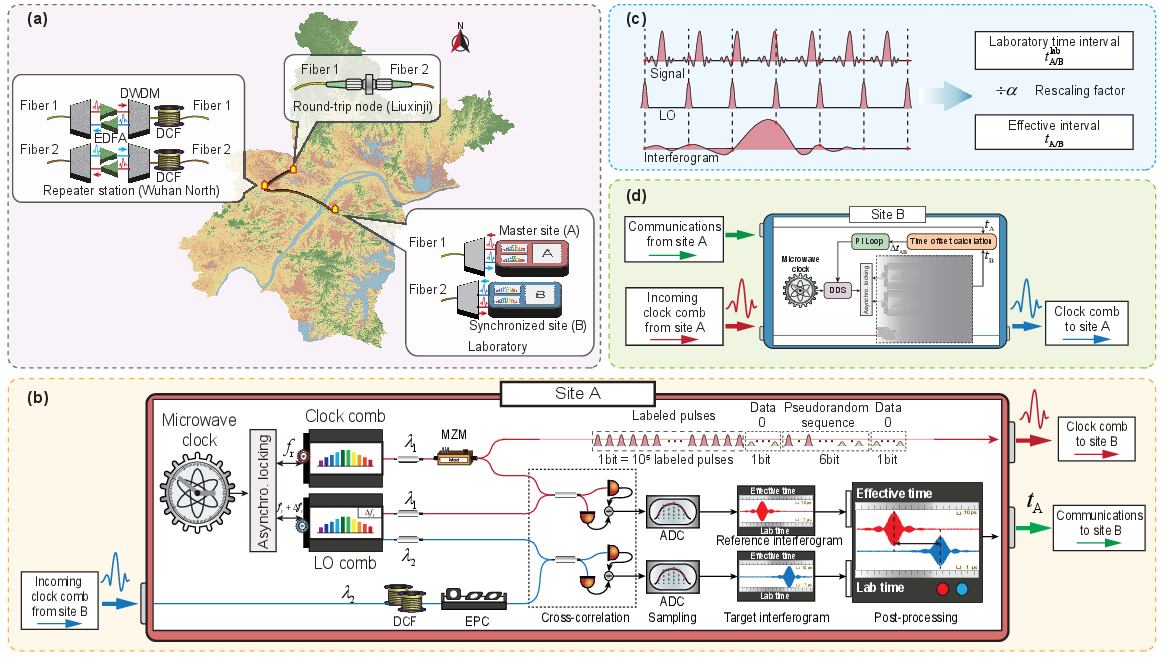}
	\caption{\textbf{Schematic of the microwave clock time synchronization.} (a) Block diagram of the experimental link. DWDM: dense wavelength division multiplexing; DCF: dispersion compensating fiber; EDFA: erbium doped fiber amplifier. (b) Detailed configuration of site A. The out-of-loop verification part is not shown here. The repetition frequencies of the clock comb and local oscillator (LO) comb has a small frequency offset $\Delta {f_{\rm{r}}} = 1$ kHz. The center frequencies of two optical filters are ${\lambda _1}$ = 1,551.72 nm and ${\lambda _2}$ = 1,550.12 nm, respectively. MZM: Mach-Zehnder modulator; EPC: electronic polarization controller; ADC: analog to-digital converter. (c) Principle of linear optical sampling. (d) Simplified configuration of site B. Site B is identical to site A, except that all of the positions of wavelengths ${\lambda _1}$ and ${\lambda _2}$ are exchanged, and a synchronization feedback system is added, which contains a real-time time offset calculation, a proportional integral (PI) loop filter and
		a direct digital synthesizer (DDS).}
	\label{link}
\end{figure*}

\section{Experimental layout}

The experiment was performed in Wuhan commercial link. As shown in Fig.~\ref{link} (a), two sites A and B are adjacent to each other and located at Wuhan laboratory. Site A and site B were connected through two geographically folded optical fibers linked in Liuxinji (round-trip node), and the total link length was 205.86 km (the total link loss was 57.18 dB). The long-haul fiber transmission necessitated the installation of two optical repeater stations at Wuhan North, each of which contained an erbium-doped fiber amplifier (EDFA) and a DCF along the terrestrial link to recover the signal impairments caused by loss and dispersion. OTWTT was achieved by using the DWDM technology over the shared fiber link.

\begin{figure*}[t!]
	\centering
	\includegraphics[width=0.9 \linewidth]{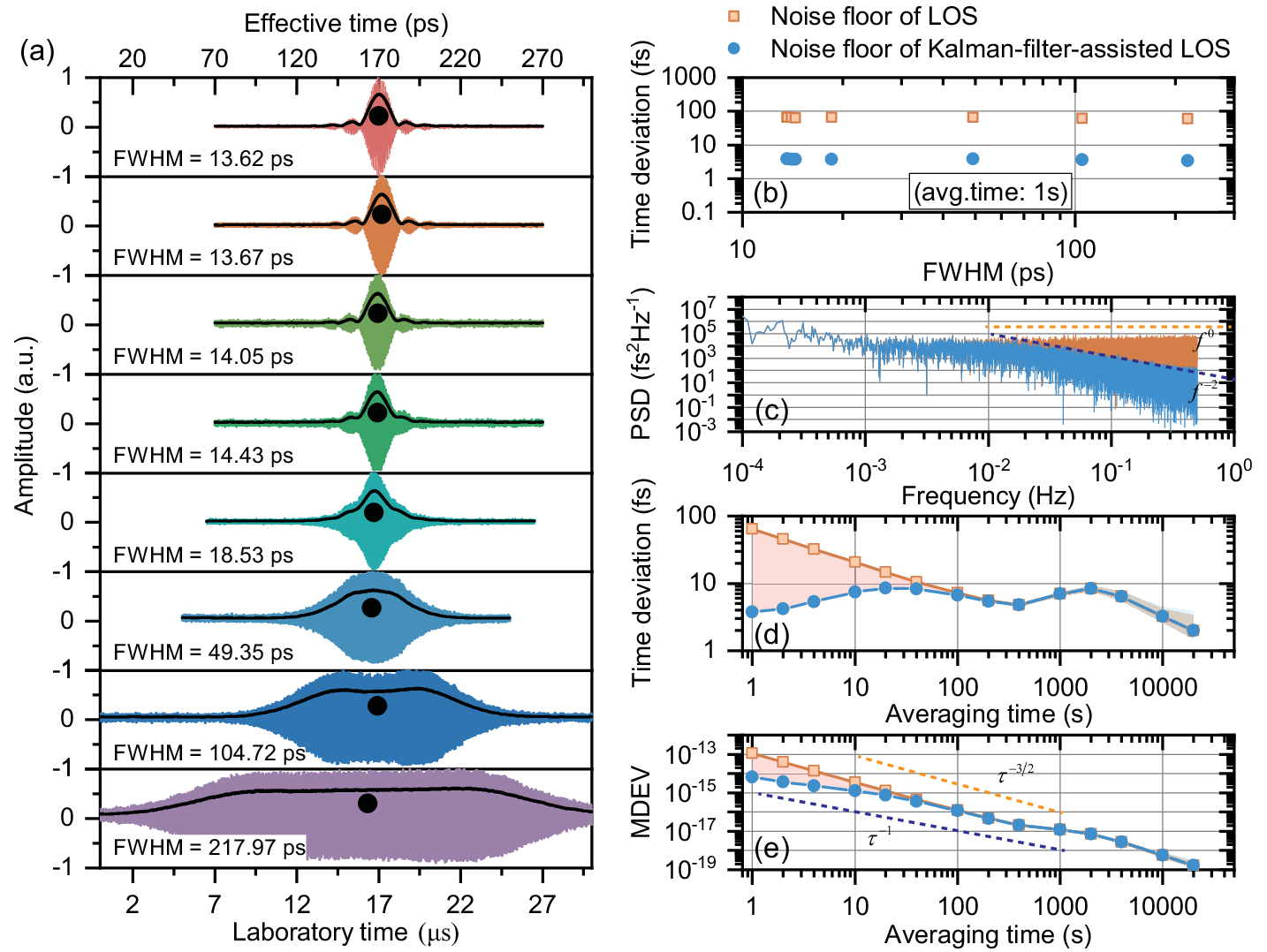}
	\caption{\textbf{Noise floor of the LOS measurement.} (a) Centroid-finding process of different interferograms by locally changing the length of the DCF. FWHM: full width at half maximum. (b) Time deviation of the noise floor over an averaging time of 1 s, calculated by the centroids, under different pulse FWHMs. (c) Power spectral densities (PSDs) before and after Kalman filtering. (d) Time deviation of the noise floor. The shaded region indicates the short-term noise suppression by the Kalman filter. (e) Modified Allan deviation (MDEV) of the noise floor. In both plots, the orange squares or solid lines indicate the noise floor of the LOS; the blue circles or solid lines indicate the noise floor of the Kalman filter–assisted LOS. The dashed lines highlight the main contributed noise types with typical slopes. The error bars indicate the 68$\%$ confidence intervals.}
	\label{Local_Results}
\end{figure*}

At each site, we used a dual comb–based LOS measurement to replace the TIC in the measurement of pulses generated by microwave clocks. As shown in Fig.~\ref{link} (b), at each site, we used two combs with the repetition frequency of ${f_{\rm{r}}} =$ 100 MHz and ${f_{\rm{r}}}+ \Delta {f_{\rm{r}}}=$ 100.001 MHz for performing LOS. The principle of LOS is shown in Fig.~\ref{link} (c). Phase locking was achieved with a locking-induced noise of 45 fs (time deviation at 1 s) through an asynchronous locking system (see Methods) to create the local timescales. 

\begin{figure*}[t!]
	\centering
	\includegraphics[width=0.9 \linewidth]{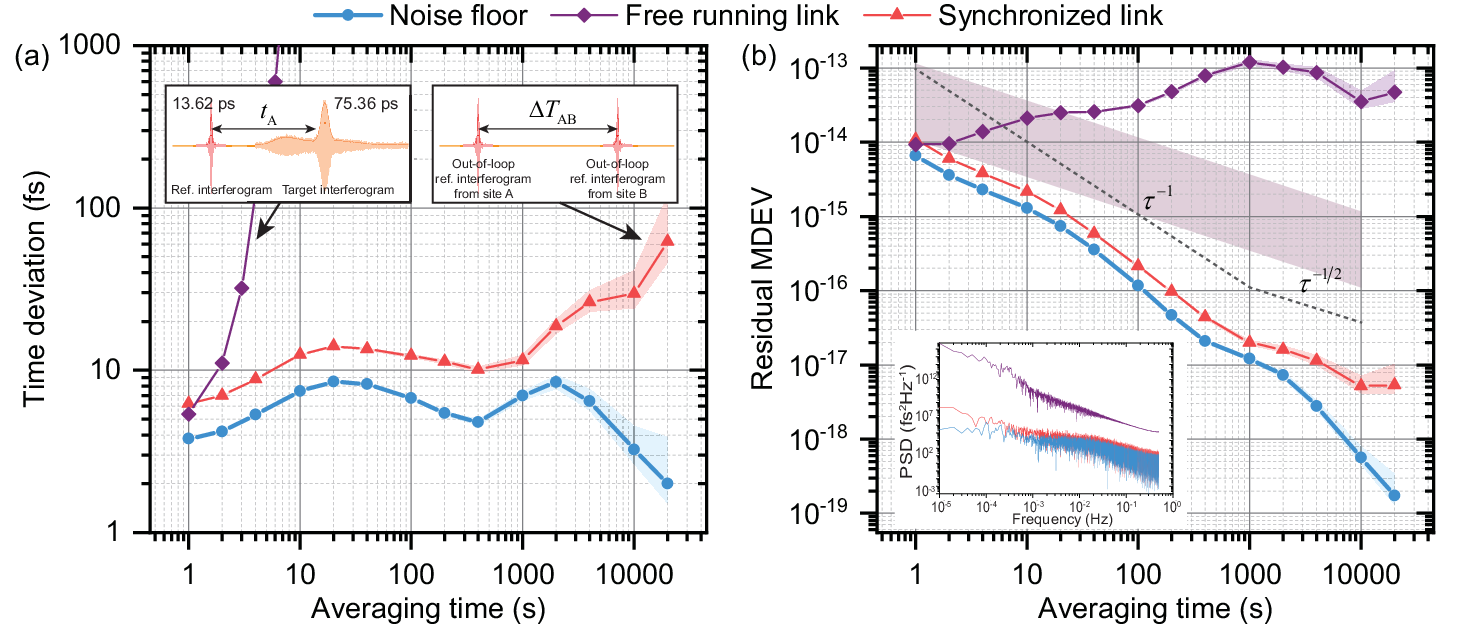}
	\caption{\textbf{Results of the dual-comb-enhanced microwave clock synchronization.} (a) Time deviations of the LOS noise floor (blue circles), one-way free-running link (purple diamonds) and synchronized link (red triangles). The free running performance was obtained by comparing the position of the centroids between local reference and target interferograms. The performance of the synchronized link was calculated from the out-of-loop measurement $\Delta {T_{{\rm{AB}}}}$ by comparing the position of the centroids between out-of-loop reference interferograms from two sites.
		(b) Residual MDEV of the corresponding results. The insert shows the PSDs of OTWTT. The synchronized result was well below the instabilities of state-of-the-art microwave clocks (shaded region~\cite{EFTF/IFC.F.Fang.2019,Metrologia.55.789.2018,Metrologia.51.174.2014,IEEE.Trans.Ultrason.Ferroelect.Freq.Contr.59.391.2012}). The dashed lines highlight the main contributed noise types with typical slopes. The error bars indicate the 68$\%$ confidence intervals.}
	\label{Synchronized_Results}
\end{figure*}

Both combs had center frequencies of 1,560 nm and a 3-dB spectrum bandwidth of approximately 15 nm. The spectral effectively covered the commonly used C-band range in fiber optic communication, which is necessary for two-way transmission. Here, we directly applied the ITU DWDM module to the output of each comb to generate band-limited pulses with different spectra. We filtered the LO comb to generate two outputs with a bandwidth of 0.8 nm (100 GHz), which were centered at 1,551.72 nm (ITU-C32) and 1,550.12 nm (ITU-C34). We filtered the clock comb to generate an output centered at ITU-C32 (for site A) or ITU-C34 (for site B). 

The use of DWDM offered two advantages. First, the spectral bandwidth of the filtered comb signals were limited to stay below Nyquist condition: $ {{f_{\rm{r}}^2} \mathord{\left/	{\vphantom {{f_{\rm{r}}^2} {\left( {2\Delta {f_{\rm{r}}}} \right)}}} \right.\kern-\nulldelimiterspace} {\left( {2\Delta {f_{\rm{r}}}} \right)}}$~\cite{Nature.Photon.7.434.2013}. Second, combs with a constrained optical bandwidth could better propagate through fibers to decrease the dispersion and nonlinear effects~\cite{Laser.and.Photonics.Reviews.16.2200167}. However, this behavior increases the uncertainty of timing discrimination compared to the free-space scenario with narrower pulses (where the typical pulse width is 0.1-1 ps)~\cite{Phys.Rev.X.6.021016.2016}.

At each site, two cross-correlations were generated. The cross-correlations were generated with the equivalent sampling rate ${f_{\rm{s}}} = {{f_{\rm{r}}^2} \mathord{\left/	{\vphantom {{f_{\rm{r}}^2} {\Delta {f_{\rm{r}}}}}} \right.	\kern-\nulldelimiterspace} {\Delta {f_{\rm{r}}}}} = 10$ THz in the optical frequency domain and the updating rate of ${T_{{\rm{update}}}} = {1 \mathord{\left/{\vphantom {1 {\Delta {f_{\rm{r}}}}}} \right.	\kern-\nulldelimiterspace} {\Delta {f_{\rm{r}}}}} =1$ ms in the laboratory timescale. The equivalent timescale stretching parameter was obtained as $\alpha = {{{f_{\rm{r}}}} \mathord{\left/{\vphantom {{{f_{\rm{r}}}} {\Delta {f_{\rm{r}}}}}} \right.\kern-\nulldelimiterspace} {\Delta {f_{\rm{r}}}}}{\rm{ = 1}}{{\rm{0}}^5}$. Then, we implemented an ADC with a 400-MHz sampling rate and a 10-bit resolution to digitize the train of interferograms, and then the data were transmitted to a FPGA controller for data processing.

The fundamental task of OTWTT is to recover timing information from exchanged optical pulses. Because signal distortion and noise accumulation were caused by the long-haul fiber transmission, high-precision timing discrimination was challenging. Instead of the commonly used heterodyne-voltage-to-time scaling~\cite{Nature.618.721.2023}, peak location~\cite{Nature.610.661.2022} or carrier phase~\cite{Phys.Rev.Lett.120.050801.2018} to define the arrival time in free-space scenarios, we used the centroid of the interferogram to discriminate the temporal position of the optical pulses, defined by Eq.~(\ref{centroidCalculation}).

In contrast to the signal fading from turbulence in the free-space scenario~\cite{Nature.618.721.2023,Nat.Commun.10.1819.2019}, the decrease in visibility of the target interferogram mainly attributes to the polarization disturbance in fiber, which is eliminated by the in-line electronic polarization controller. Therefore, the hold-over behavior of the Kalman filter was not considered in our experiments. In this paper, we focused on the Kalman filter to limit the amount of measurement noise~\cite{IEEE.Trans.Instrum.Meas.64.449.2015,IEEE.Trans.Ultrason.Ferroelect.Freq.Contr.52.289.2005}. We assumed two zero-mean Gaussian
random processes in the prior noise model of Kalman filter
to represent the random-walk phase noise
and random-walk frequency noise for real-time data processing~\cite{Optica.3.441.2016}, because these two noises are usually dominant in the microwave clock. This clock model is often applied to analyze the clock offset and Kalman-filter-based clock synchronization~\cite{IEEE.Trans.Instrum.Meas.64.449.2015,IEEE.Trans.Ultrason.Ferroelect.Freq.Contr.52.289.2005,IEEE.Trans.Instrumentation.Measurement.60.2011}.

Based on OTWTT, the equivalent time offset between clocks was obtained by subtracting the two time difference results as follows:
\begin{equation}
	\Delta {t_{{\rm{AB}}}} = \frac{1}{2}\left( {{t_{\rm{B}}} - {t_{\rm{A}}}} \right) + {t_{{\rm{NR}}}},
	\label{time_offset}
\end{equation}
where ${t_{{\rm{NR}}}} = {{\left( {\Delta {\tau _{{\rm{BA}}}} - \Delta {\tau _{{\rm{AB}}}}} \right)} \mathord{\left/{\vphantom {{\left( {\Delta {\tau _{{\rm{BA}}}} - \Delta {\tau _{{\rm{AB}}}}} \right)} 2}} \right.\kern-\nulldelimiterspace} 2}$ is the non-reciprocal propagation time delay between forward transmission time ${\Delta {\tau _{{\rm{AB}}}}}$ and backward transmission time ${\Delta {\tau _{{\rm{BA}}}}}$. In the experiment, ${\Delta {\tau _{{\rm{AB}}}}}$ and ${\Delta {\tau _{{\rm{BA}}}}}$ were estimated in advance based on the link length, and the results were pre-stored in FPGA. Based on the reciprocity assumption, ${t_{{\rm{NR}}}}$ is commonly set to zero. However, many scenarios such as asymmetric delay attacks in the fiber~\cite{IEEE.Trans.Instrum.Meas.72.1.2023} will break this assumption, which will be discussed later.

\section{Results}

\subsection{Noise floor of the LOS measurement}
To quantify the noise floor of the microwave clock–based LOS measurement, we replaced the incoming clock comb by the local generated clock comb at each site and calculated the time difference between reference and target interferograms. Due to the DWDM filtering, the initial width of the filtered pulse was approximately 13.6 ps in the effective time (which corresponds to 1.36 ${\mu \rm{ s}}$ in lab time). Then, we used local DCFs with different length to locally simulate the effect of the fiber transmission on the pulse shape, as shown in Fig.~\ref{Local_Results} (a). When the dispersion-induced pulse width increased, the peak became more difficult to find because the pulse flattened. By contrast, as shown in Fig.~\ref{Local_Results} (b), in the pulse width of 13.62-217.97 ps, the short-term time deviation remained stable (65 $\pm $ 2 fs for 1 s of averaging time) by the centroid estimation, which demonstrates the insensitivity to waveforms.

Figure~\ref{Local_Results} (c) shows noise power spectral densities (PSDs) of the measurement process. The effective bandwidth of the Kalman filter was 0.01 Hz, and the PSD was reduced by approximately three orders of magnitude reduction at 0.5 Hz to 1.2 ${\rm{f}}{{\rm{s}}^{\rm{2}}}{\rm{H}}{{\rm{z}}^{{\rm{ - 1}}}}$, which is close to the noise floor of the optical clock–based LOS measurement~\cite{Nature.618.721.2023,Nature.610.661.2022,Optica.3.441.2016}. In addition, Figs.~\ref{Local_Results} (d) and (e) show that the Kalman filter–assisted LOS optimizes the measurement instability for a short averaging time. The shaded region shows the noise suppression below 100 s. In Fig.~\ref{Local_Results} (d), the time deviation of noise floor was optimized to 3.7 $\pm $ 0.1 fs for 1 s of averaging time, which can support the measurement of femtosecond-level fiber changes. In Fig.~\ref{Local_Results}~(e), the modified Allan deviation (MDEV) changed the slope of the noise floor from ${\tau ^{ - {3 \mathord{\left/	{\vphantom {3 2}} \right.\kern-\nulldelimiterspace} 2}}}$ to ${\tau ^{ - 1}}$ after the Kalman filtering ($\tau$ is the averaging time in seconds). Thus, the dominant white phase measurement noise was suppressed to the lower flicker phase noise region at short timescales (below 100 s), which determined the noise limitation of the time offset measurement.

\subsection{Free-running link}

We then used our system to investigate the one-way free-running link. Specifically, we set both sites that shared a common master clock to test the open-loop link noise from A to B. After passing through the link, the incoming clock comb accumulated link effects, which resulted in the pulse impairments. One impairment is dispersion, which caused pulse broadening and distortion. The results show both second-order dispersion, which caused pulse broadening, and third-order dispersion, which resulted in pulse asymmetry and distortion. Additionally, the link noise and amplified spontaneous emission (ASE) noise introduced by EDFAs inevitably introduced intensity fluctuations in the optical pulses. Both dispersion-induced noise and ASE-induced intensity fluctuations decreased the pulse coherence, which ultimately affected the measurement jitter of the optical cross-correlation.

For the dispersion compensation, even if we use DCFs at the repeater station, a complete dispersion compensation for ultrashort pulse remains challenging. Therefore, we used a small amount of DCF at the receiver site to compensate for the residual dispersion in a fine-tuning manner. The cross-correlation after compensation was provided by the target interferogram in the insert of Fig.~\ref{Synchronized_Results} (a) with a pulse width of 75.36 ps, which would not cause significant measurement uncertainties due to pulse distortion. For more precise dispersion compensation, techniques such as liquid crystal modulation may be used~\cite{Opt.Lett.23.283.1998}. The effect of the intensity fluctuations was suppressed by the centroid estimation to smooth the intensity noise. When we calculated the centroid, as shown in Fig.~\ref{Synchronized_Results}~(a), the time deviation introduced by the link was 5 fs at 1 s, which was optimized by an order of magnitude compared to the direct peak-location finding method in our experiment.

Imperfect dispersion and nonlinearities led to waveform distortion, which mainly affected the estimation of time offset in two aspects. First, there was a fixed systematic offset relative to the ideal position of centroid, which guided us to strictly calibrate the offset after the imperfect compensation. Second, these could lead to insufficient concentration of signal energy near the centroid, which could result in larger fluctuations when calculating the envelope and centroid using limited data. Therefore, it is important to carefully select both the amount of data and the data region used for calculating the centroid when the compensation was not perfect. We will leave the investigation of the impact of imperfect dispersion compensation and nonlinearities on system performance for future work.

\subsection{Synchronized link} After fully characterizing the 205.86-km fiber link, by performing an OTWTT to cancel the link's influence and closing the synchronized feedback loop, we obtained the performance of the synchronized link, which was continuously acquired for over 24 hours. The synchronized link in our work, distinctly from the active stabilization link mentioned in Ref.~\cite{Light.Sci.Appl.23.2016}, refers to the pulse-synchronization link achieved after eliminating time deviations through feedback systems. As shown in Fig.~\ref{link} (d), the time offset calculation, which acted as the error signal, was fed into the digital feedback system that contained a proportional integral loop filter and
a direct digital synthesizer (DDS). It also had a 30-Hz synchronization bandwidth to force the minimized residual timing jitter of the clock output to synchronize two microwave clocks. Note that the synchronization was actually achieved between the repetition frequency of the remote clock comb and the local clock comb, with each piece of time information carried from its local clock, respectively.

\begin{figure*}[t]
	\centering
	\includegraphics[width=0.9 \linewidth]{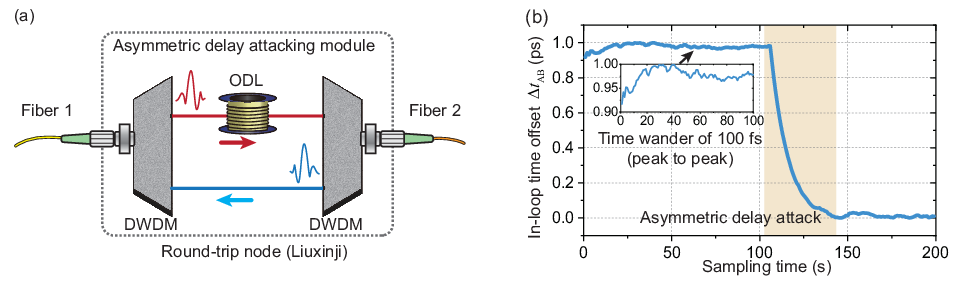}
	\caption{\textbf{Result of the asymmetric delay attack.} (a) Asymmetric delay attacking module at the round-trip node. Two DWDMs were used to separate the forward and backward optical pluses, and an optical delay line (ODL) in the forward path was used to adjust the reciprocal time delays between forward and backward transmissions. (b) The in-loop time offset changed when the sub-picosecond-level asymmetric delay attack was used. The peak-to-peak time wander was approximately 100 fs without attack.}
	\label{Delay attack}
\end{figure*}

Fig.~\ref{Synchronized_Results} (a) shows a time deviation. The starting point was the floor at approximately 6.23 fs and reached another floor of 10 fs at 400 s, which contributes to the short-term noise suppression by Kalman filter–assisted LOS. The trend is consistent with the measurement noise floor for an averaging time of up to 1,000 s, which indicates that the residual synchronized noise strongly depends on the type of noise in the measurement part and the degree of noise. The synchronized link exhibits slightly poorer stability for very short averaging time than the free-running link. The main reason is the non-reciprocity in the bidirectional amplifiers at the repeater station, which makes it difficult to cancel the non-common-mode ASE noise. The slope dramatically changed beyond 1,000 s, which shows that the time deviation is limited by the thermal effect for longer averaging times. This issue will be improved by an active thermal stabilization of experimental components.

The residual MDEV between remote parties show the performance of the time-frequency transfer in Fig.~\ref{Synchronized_Results} (b). The MDEV we provide is not the stability of a single microwave clock, but two clocks' relative stability after synchronization. The relative stability between two rubidium clocks is $5.5 \times {10^{ - 13}}$ at 1 s and reaches the floor of $8.56 \times {10^{ - 14}}$ at 2,000 s without synchronization. The residual MDEV began at $1 \times {10^{ - 14}}$ at 1 s and reached below $6 \times {10^{ - 18}}$ at 10,000 s. The residual MDEV was below the state-of-the-art
microwave clocks~\cite{EFTF/IFC.F.Fang.2019,Metrologia.55.789.2018,Metrologia.51.174.2014,IEEE.Trans.Ultrason.Ferroelect.Freq.Contr.59.391.2012} after 1 s of averaging time and more than two orders of magnitude better after the long averaging time. The slope was ${\tau ^{ - 1}}$ below 1,000 s, which indicates that the dominant noise at short timescales is flicker phase noise, and the residual MDEV follows ${\tau ^{ - 1/2}}$ power laws after 1,000 s averaging time. All of the results are consistent with the time deviation.

We also compared the stability result of $1 \times {10^{ - 14}}{\tau ^{ - 1}}$ in our experiment with the free-space scenarios that synchronized the optical clocks.
The stability was approximately one order of magnitude less than that of Ref.~\cite{Nature.610.661.2022}, given by a few $ {10^{ - 15}}{\tau ^{ - 1}}$. This lower starting point was mainly due to the higher noise floor of the LOS measurement and ASE noise of the in-line EDFAs, which relatively weakened the coherence of optical pulses and caused a larger timing jitter in the microwave-clock-synchronization scenarios than in the optical oscillator–based scenarios. Moreover, compare to the result of $2.8 \times {10^{ - 15}}{\tau ^{ - {3 \mathord{\left/{\vphantom {3 2}} \right.\kern-\nulldelimiterspace} 2}}}$ in Ref.~\cite{Nature.618.721.2023}, the trend of ${\tau ^{ - 1}}$ comes from the flicker phase noise caused by the technical noise of devices, so the delay-unsuppressed white-phase noise especially with low frequency~\cite{J.Opt.Soc.Am.B.25.1284.2008,Opt.Lett.32.3056.2007} could not be detected. Thus, we should optimize the microwave clock–based LOS measurement. Nevertheless, this result, which is close to the optical clock time synchronization, greatly closes the gap of microwave-microwave time synchronization in the entire time-frequency network.

Recently, similarly to what we did, H. J. Kang\textit{ et al.}~\cite{Nat.Commun.10.4438.2019} utilized the filtering comb with WDMs to transmit multiple optical carriers in free space scenarios, which also provides a new possibility for comb-based time-frequency transfer.

\subsection{Probing fiber delay non-reciprocity} Similar to the probe of non-reciprocity from multi-path effects~\cite{Nature.618.721.2023} or motion platforms~\cite{Nat.Commun.10.1819.2019} in free space, many impacts can affect the non-reciprocity in fiber-based OTWTT, which affects the true clock-offset estimation results. A particular example of the recently extensively concerned potential asymmetric delay attacks~\cite{IEEE.Trans.Instrum.Meas.72.1.2023,Applied.Physics.Letters.115.141101.2019,IEEE.J.Sel.Top.Signal.Process.12.749.2018}. The improvement of the clock-offset measurement enables us to sense the subtle delay non-reciprocity of the fiber link and improves the security of long-haul fiber transmission.

To verify this claim, we used an asymmetric delay attacking module at the round-trip node to simulate the potential invasive attack, as shown in Fig.~\ref{Delay attack} (a). Two DWDMs were used to separate the forward and backward optical pluses. An optical delay line (ODL) in the forward path was exploited to adjust the relative time delays between forward and backward transmissions, and the reciprocity was breakdown. As we discussed, the non-reciprocal propagation time delay ${t_{{\rm{NR}}}}$ in Eq.~(\ref{time_offset}) was set to zero before we started the synchronization. If the non-reciprocal traveling time was stable, the calculated time offset $\Delta {t_{{\rm{AB}}}}$ only reflected the offset between the two clocks (because the link drift was canceled by the OTWTT method), which can be compensated on the remote clock through closed-loop feedback process. However, the term ${t_{{\rm{NR}}}}$ appeared when we adjusted the ODL to perform asymmetric delay attacks, which made the estimated clock offset deviate from the true value. 

Here, we only set a common clock of two sites to probe the link variation to show the ability of pulse measurement in the fiber and ignored the clock-offset drift. When there is a clock offset, the link change can also be obtained by establishing a clock-offset dynamics model to eliminate the influence of the clock~\cite{IEEE.Trans.Instrum.Meas.72.1.2023}. As Fig.~\ref{Delay attack} (b) shows, the in-loop time offset $\Delta {t_{{\rm{AB}}}}$ suddenly changed when we applied a sub-picosecond-level delay on one path, which caused a mismatch between synchronized clocks and compromised the security. This type of asymmetric delay attack cannot be identified and exceed the noise level of the measurement instruments. The time wander of the system was approximately 100 fs in our system, which limits the precision of probing fiber delay slips. This process can be optimized by precisely controlling the temperature of laboratory devices.

This result indicates that the transmission of microwave clocks can identify delay variations caused by sub-ps-level attacks, which is three orders of magnitude better than the reported TIC scheme with the sub-nanosecond precision~\cite{IEEE.Trans.Instrum.Meas.72.1.2023}. This significant improvement helps us precisely identify the sudden phase changes caused by unexpected factors in the link, which can more securely evaluate the clock offset.

\section{Conclusion}

In summary, we have demonstrated a femtosecond-level time synchronization
between remote microwave clock–based timescales over a commercial fiber link of 205.86 km. The residual time deviation after synchronization was 6.23 fs at 1 s, and the instability was below $6 \times {10^{ - 18}}$ for an averaging time longer than 10,000 s with the dominant noise from the microwave clock–based LOS measurement. This configuration can make the high time-resolution feature of the optical frequency comb compatible with most existing fiber infrastructures, which greatly promotes the application range of fiber-based time-frequency networks. This result fully merges the bottleneck of time synchronization based on microwave clocks in a hybrid time-frequency network to achieving the interconnection between any microwave and optical clocks. Thus, this method has the potential for microwave-based precise applications such as positioning, navigation, timing and coherent sensing. Moreover, this technique enhances the probe ability of subtle variations of fiber, which will generate new possibilities
for channel modeling and a secure time-transfer network to resist potential delay attacks. An interesting extension of this work will further investigate the quantum-limited properties~\cite{Nature.618.721.2023} in the microwave clock–based fiber scenarios.

\section{Methods}

\subsection{Asynchronous locking system}
In our experiment, the repetition frequencies of two combs were phase-locked to a microwave clock to guarantee the coherence of the pulse-envelope phase, while the CEO frequency was free running. The reason is that the CEO frequency has negligible influence on the envelope measurement at an appropriate sampling rate~\cite{Photon.Res.11.2222.2023}. 

To maintain the pulse-envelope phase synchronization between two combs with a small frequency offset, we developed an asynchronous locking system, which contained a high-precision single-input two-output DDS, and two fiber-loop optical-microwave phase detectors
(FLOM-PDs)~\cite{Opt.Lett.37.2958.2012}. The single-input two-output DDS was used to generate two electrical signals with repetition frequencies of 1 GHz and 1.00001 GHz, which were synchronized to the 10-MHz microwave clock and served as the references of two combs. To maintain the synchronization between the outputs of the DDS and two combs, we locked the repetition frequencies of two combs to the outputs of the DDS using the FLOM-PD, where the 10-th harmonic of the respective combs were used to improve the locking stability. The overall locking-induced time deviation of the system was 45 fs at 1 s, and the corresponding phase noise at typical frequency points were -55 ${\rm{dBc/Hz}}@{10^{ - 3}}{\rm{Hz}}$, -100 ${\rm{dBc/Hz}}@{10^{ - 1}}{\rm{Hz}}$, -123 ${\rm{dBc/Hz}}@{10}{\rm{Hz}}$, and the floor of -130 ${\rm{dBc/Hz}}@{10^{ 5}}{\rm{Hz}}$ was obtained. Future improvements in the FLOM-PD performance can be attained through the use of high-order harmonics for locking and optimizing the noise characteristics of electronic devices.

\subsection{Elimination of the ambiguity}
Instead of the modulated laser scheme to remove the ambiguity range, we used the MZM to directly encode time markers on the intensity of the optical pulses. The MZM was driven by the digital-to-analog converters (DAC) synchronized to the local clock. High control voltage was supplied to suppress the intensity of the optical pulse to label data ``0", whereas no control voltage was supplied to maintain the intensity of the optical pulses to label data ``1". 

Since we collected interferograms with a refresh rate of 1 kHz, the modulated frequency of MZM was set to 1 kHz to label symbols on the initial part of the interferograms of every second to represent the timestamp. Therefore, one labeled interferogram corresponds to $10^5$ labeled optical pulses. Here, the timestamp was encoded by a pseudorandom binary sequence (PRBS) with 10 bits (corresponding to 10 ms duration) per second. The frame structure of the PRBS is as follows:
The first and last bits were fixed as ``0", and the middle 8 bits used pseudo-random numbers to simulate the timestamps.

The next step was to recognize the frames. First, once the measurement voltage of the reference interferogram was 0, the frame header recognition began. Second, at each site, after identifying all pseudo-random numbers and receiving the last ``0" symbol, we labeled the next interferogram as the first pulse and recorded the time on the FPGA. The target interferogram was similarly measured. The synchronization was maintained until the symbol ``0" of the next frame was received, and the measurement time difference was maintained during the frame recognition process.

\subsection{Out-of-loop verification}
To verify the time offset of the clocks after synchronization, the out-of-loop measurement of the time offset $\Delta {T_{{\rm{AB}}}}$, which is independent
from the in-loop calculated time offset $\Delta {t_{_{{\rm{AB}}}}}$, is necessary. Here, we also filtered an output on both combs at each site to generate the out-of-loop reference interferogram that is independent from the in-loop signals to verify the synchronization results. The out-of-loop reference interferograms at two sites contained the timing information of two clocks and could directly eliminate uncertainty from the photoelectronic detection. Then, the out-of-loop verification was performed by measuring the relative time interval of two out-of-loop reference interferograms from two sites using an ADC and transmitted to a computer with a real-time data refresh rate of 1 s, which is independent of the FPGA processor used for synchronization. Therefore, the PSD is available only for Fourier frequencies below 0.5 Hz. The limited sampling rate hampers the acquisition of PSD spectra at high Fourier frequencies, thereby posing difficulties in estimating MDEVs using the integration method described in Refs.~\cite{1275096,84972}. In the future, we hope to increase the refresh rate of the our out-of-loop verification system to obtain more complete PSD spectrum information.

	\section{Funding} This work was supported by the National Natural Science Foundation of China (Grants Nos. 62201012), and the National Hi-Tech Research and Development (863) Program.
	

	\section{Author contributions}
	Z.C., D.Y., G.L. and Y.Z. performed the experiment and overall data
	analysis. Z.C., D.Y. and Y.Z. performed the theoretical analysis. Z.C., D.Y., and G.L. processed the data. Z.C. and H.G. contributed to all parts of the work. Z.C. and D.Y. conceived the experiment, and Z.C., S.Y. B.L. and H.G. supervised the project. All of the authors were involved in the discussion and interpretations of the results. Z.C. wrote the manuscript, and D.Y., G.L., Y.Z., S.Y., B.L. and H.G. provided revisions. 
	
	\section{Disclosures} The authors declare no conflicts of interest.

	\section{Data Availability} Data supporting the results presented in this paper are available from the authors upon reasonable request.


\section{Acknowledgments}

This work was supported by the National Natural Science Foundation of China (Grants Nos. 62201012), and the National Hi-Tech Research and Development (863) Program.

\section{AUTHOR CONTRIBUTIONS}
Z.C., D.Y., G.L. and Y.Z. performed the experiment and overall data
analysis. Z.C., D.Y. and Y.Z. performed the theoretical analysis. Z.C., D.Y., and G.L. processed the data. Z.C. and H.G. contributed to all parts of the work. Z.C. and D.Y. conceived the experiment, and Z.C., S.Y. B.L. and H.G. supervised the project. All of the authors were involved in the discussion and interpretations of the results. Z.C. wrote the manuscript, and D.Y., G.L., Y.Z., S.Y., B.L. and H.G. provided revisions.

\section{COMPETING INTERESTS}
The authors declare no competing interests.

\end{document}